\documentclass[11pt]{article}

\usepackage{amsmath,amssymb,bm}
\usepackage{hyperref}
\usepackage{authblk}

\numberwithin{equation}{section}

\newcommand{\Tr}{\mathrm{Tr}}

\newcommand{\cL}{\mathcal{L}}
\newcommand{\cH}{\mathcal{H}}
\newcommand{\eps}{\varepsilon}

\newcommand{\Pl}{\mathrm{Pl}}
\newcommand{\LP}{L_{\Pl}}
\newcommand{\tauP}{\tau_{\Pl}}

\newcommand{\gr}{\eps}

\newcommand{\Hsa}{\cH_{\mathrm{sa}}}
\newcommand{\Hasa}{\cH_{\mathrm{asa}}}

\title{In models of spontaneous wave-function collapse, why only fermions collapse, not bosons?}
\author{Tejinder P.\ Singh}
\affil{\it Tata Institute of Fundamental Research, Homi Bhabha Road, \\ {\it Mumbai 400005, India}}
\date{}

\begin{document}
\maketitle

\begin{abstract}
\noindent Objective collapse models are often implemented so that collapse acts only on the
fermionic (matter) sector, while bosonic fields do not undergo fundamental collapse.
In generalized trace dynamics (GTD), spontaneous localization is expected to arise
when the trace Hamiltonian has a significant anti-self-adjoint component.
In this note we show, starting from the STM-atom (spacetime-matter atom) trace Lagrangian written in terms of
two inequivalent matrix velocities $\dot Q_1$ and $\dot Q_2$, that the purely bosonic
subsector admits a self-adjoint Hamiltonian, whereas the fermionic sector carries an
intrinsic anti-self-adjoint contribution. The key structural input is that making the
trace Lagrangian bosonic requires insertion of two \emph{unequal} odd-grade Grassmann
elements $\beta_1\neq \beta_2$. Assuming natural adjoint properties for these elements,
we compute the trace Hamiltonian explicitly via trace-derivative canonical momenta
(with bosonic and fermionic variations treated separately) and isolate the resulting
anti-self-adjoint term. This provides a first-principles mechanism, within GTD, for
why only fermionic degrees of freedom act as collapse channels.
\end{abstract}

\section{Introduction}

% --- Add the following opening paragraphs at the start of the Introduction,
%     immediately BEFORE the current first paragraph on spontaneous collapse. ---

A recurring conceptual obstacle in quantum gravity is the role of time: standard quantum
theory (and, in particular, quantum field theory) is formulated with respect to an external
classical time parameter, whereas general relativity treats spacetime geometry as dynamical.
Generalized trace dynamics (GTD) \cite{Singh2024TraceDynamicsE8xE8}   addresses this tension by building on Adler's \emph{trace dynamics}
(TD) \cite{adler-trace}, a deterministic matrix-valued Lagrangian/Hamiltonian dynamics in which the action is the
trace of a polynomial in noncommuting matrix degrees of freedom (bosonic even-grade matrices and
fermionic odd-grade matrices). A key structural feature of TD is its global unitary invariance,
which gives rise to a novel conserved quantity---the Adler--Millard charge---with dimensions of action,
\begin{equation}
\tilde C \;=\; \sum_{r\in B}\,[q_r,p_r]\;-\;\sum_{r\in F}\,\{q_r,p_r\}\,,
\end{equation}
absent in ordinary classical dynamics. When the statistical mechanics of TD is developed, equipartition
of $\tilde C$ at equilibrium yields the canonical (anti)commutation relations and an emergent unitary
quantum (field) dynamics for the canonical averages. Conversely, when the anti-self-adjoint part of
the fundamental trace Hamiltonian is not negligible, fluctuations about equilibrium can drive effective
nonunitary nonlinear stochastic dynamics and spontaneous localization \cite{KakadeSinghSingh2023SpontaneousLocalisation}.

GTD extends this pre-quantum framework toward quantum gravity by promoting gravitational degrees of
freedom to matrices and by replacing classical spacetime labels by a noncommutative ``pre-spacetime''
(quaternionic/octonionic, in the models of interest), thereby aiming for a formulation of quantum dynamics
without reference to an external classical time (evolution may be parametrized by Connes time $\tau$) \cite{Singh2024TraceDynamicsE8xE8}.
The fundamental entities are \emph{atoms of spacetime-matter} (STM atoms): matrix degrees of freedom
$q_i=q_{B i}+q_{F i}$ whose bosonic components are interpreted as ``atoms of spacetime,'' while the full
STM atom represents a fermion together with the bosonic fields it sources (for example, an electron together
with its electromagnetic, weak, and gravitational fields). Entanglement among many STM atoms, followed by
spontaneous localization, is conjectured to yield classical spacetime geometry and classical macroscopic
bodies \cite{Singh:2018cwg}. In this setting, objective collapse is expected precisely in regimes where the fundamental trace
Hamiltonian develops a significant anti-self-adjoint component. This motivates the specific structural
question addressed below: whether the STM-atom trace Lagrangian underlying GTD generates an intrinsic
anti-self-adjoint contribution only in the fermionic sector---thereby explaining why collapse models
effectively act on fermions and not on bosonic degrees of freedom.

% --- Existing opening paragraph on spontaneous collapse follows here. ---

Spontaneous collapse models (e.g.\ CSL-type modifications of quantum dynamics) \cite{PhysRevD.34.470, PhysRevA.42.78, PhysRevA.40.1165, Bassi:2003gd, collreview}
are commonly set up so that the collapse-inducing nonunitarity couples to matter degrees
of freedom, while bosonic field degrees do not constitute independent collapse channels.
From the viewpoint of generalized trace dynamics (GTD), a natural structural origin of
collapse is the appearance of an anti-self-adjoint (ASA) component of the trace Hamiltonian:
when ASA effects are significant, coarse-graining can yield effective nonunitary stochastic
dynamics and spontaneous localization (as in Adler-type trace dynamics arguments \cite{adler-trace}).

A central open issue for GTD phenomenology is then:
\begin{quote}
\emph{Does the fundamental GTD STM-atom Lagrangian generate ASA contributions only
in the fermionic sector, thereby explaining why only fermions collapse?}
\end{quote}

In this note we answer this question in the affirmative, for the STM-atom action written
in terms of two inequivalent matrix degrees of freedom $Q_1$ and $Q_2$.
We work directly at the level of the trace Lagrangian and its Legendre transform,
using trace derivatives in the sense of Adler, and we keep variations with respect to
bosonic and fermionic variables strictly separated.

\section{STM-atom action and unified variables}

\subsection{Action}

We begin with the unified  STM-action \cite{Singh2024TraceDynamicsE8xE8}
\begin{equation}
\frac{S}{\hbar}
=
\frac12 \int \frac{d\tau}{\tauP}\;
\Tr\!\left[
\lambda\,\dot Q_1^{\dagger}\,\dot Q_2
\right],
\qquad
\lambda := \frac{\LP^2}{L^2}.
\label{eq:action}
\end{equation}
Here $\tau$ is Connes time and $\dagger$ denotes the GTD adjoint on matrices with
Grassmann entries.

The unified velocities are defined (with bosonic/fermionic splitting) by
\begin{align}
\dot Q_B &= \frac1L\left(i\alpha\,q_B + L\dot q_B\right),
&
\dot Q_F &= \frac1L\left(i\alpha\,q_F + L\dot q_F\right),
\label{eq:QBQFdefs}
\\[4pt]
\dot Q_1^\dagger &= \dot Q_B^\dagger + \lambda\,\beta_1\,\dot Q_F^\dagger,
&
\dot Q_2 &= \dot Q_B + \lambda\,\beta_2\,\dot Q_F,
\label{eq:Q1Q2defs}
\end{align}
where $\beta_1,\beta_2$ are \emph{odd-grade} Grassmann elements inserted so that the
trace Lagrangian is bosonic. The dynamics requires $\beta_1\neq \beta_2$ \cite{Maithresh:2019ntp}. Also, $\alpha$ is a real number which stands for the Yang-Mills coupling constant.

\subsection{Grading, adjoint, and graded cyclicity}

We write $\gr(X)\in\{0,1\}$ for the Grassmann parity:
$\gr(X)=0$ for bosonic (even-grade) matrices and $\gr(X)=1$ for fermionic (odd-grade) matrices.
We assume
\begin{equation}
\gr(\dot Q_B)=0,\qquad \gr(\dot Q_F)=1,\qquad \gr(\beta_a)=1\quad(a=1,2).
\end{equation}
Thus $\beta_a \dot Q_F$ and $\beta_a \dot Q_F^\dagger$ are even, so $\dot Q_1^\dagger$ and $\dot Q_2$
are bosonic as required by the form of the trace action.

For matrices with Grassmann entries one uses the standard involution
\begin{equation}
(AB)^\dagger = B^\dagger A^\dagger,\qquad \Tr(A)^\dagger = \Tr(A^\dagger),
\label{eq:adjoint}
\end{equation}
together with graded cyclicity of the trace for homogeneous factors:
\begin{equation}
\Tr(AB) = (-1)^{\gr(A)\gr(B)}\,\Tr(BA).
\label{eq:gradedcyclic}
\end{equation}
(Equivalently, a cyclic shift of a homogeneous factor $X$ through a product $Y$
yields $\Tr(YX)=(-1)^{\gr(X)\gr(Y)}\Tr(XY)$.)

\section{Trace-derivative canonical momenta and Hamiltonian}

\subsection{Trace Lagrangian}

From \eqref{eq:action} the trace Lagrangian density (in $\tau$) is
\begin{equation}
\cL
=
\frac{\hbar}{2\tauP}\;
\Tr\!\left[\lambda\,\dot Q_1^\dagger \dot Q_2\right].
\label{eq:Lagrangian}
\end{equation}
Crucially, $\cL$ depends on the bosonic velocities $\dot Q_B,\dot Q_B^\dagger$ and
fermionic velocities $\dot Q_F,\dot Q_F^\dagger$ only through \eqref{eq:Q1Q2defs}.

\subsection{Trace derivatives (bosons and fermions varied separately)}

Following Adler, the trace derivative is defined by placing the variation to the far right:
\begin{equation}
\delta \cL = \Tr\!\left(\frac{\delta \cL}{\delta O}\,\delta O\right),
\label{eq:tracederivative}
\end{equation}
where for fermionic $O$ the graded cyclicity \eqref{eq:gradedcyclic} is used to move $\delta O$
to the right, generating sign factors only when $\delta O$ passes odd-grade factors.
In the present computation we \emph{do not} vary with respect to mixed variables;
we only vary with respect to $\dot Q_B, \dot Q_B^\dagger$ (bosonic) and
$\dot Q_F, \dot Q_F^\dagger$ (fermionic).

\subsection{Canonical momenta}

Define the overall prefactor
\begin{equation}
c := \frac{\hbar}{2\tauP}.
\end{equation}

\paragraph{Bosonic momenta.}
Varying \eqref{eq:Lagrangian} with respect to $\dot Q_B$ and $\dot Q_B^\dagger$ gives
\begin{align}
\Pi_B &:= \frac{\delta \cL}{\delta \dot Q_B}
= c\,\lambda\,\dot Q_1^\dagger,
\label{eq:PiB}
\\
\Pi_B^\dagger &:= \frac{\delta \cL}{\delta \dot Q_B^\dagger}
= c\,\lambda\,\dot Q_2.
\label{eq:PiBdag}
\end{align}

\paragraph{Fermionic momenta.}
Using $\dot Q_2=\dot Q_B+\lambda\beta_2\dot Q_F$,
\begin{equation}
\Pi_F := \frac{\delta \cL}{\delta \dot Q_F}
= c\,\lambda\,\dot Q_1^\dagger\cdot(\lambda\beta_2)
= c\,\lambda^2\,\dot Q_1^\dagger\,\beta_2.
\label{eq:PiF}
\end{equation}
Similarly, using $\dot Q_1^\dagger=\dot Q_B^\dagger+\lambda\beta_1\dot Q_F^\dagger$ and
bringing $\delta\dot Q_F^\dagger$ to the right inside the trace using \eqref{eq:gradedcyclic},
\begin{equation}
\Pi_F^\dagger := \frac{\delta \cL}{\delta \dot Q_F^\dagger}
= c\,\lambda^2\,\dot Q_2\,\beta_1.
\label{eq:PiFdag}
\end{equation}

Equations \eqref{eq:PiB}--\eqref{eq:PiFdag} are the separated bosonic/fermionic canonical
momenta required for a trace-dynamics Legendre transform.

\subsection{Legendre transform and explicit Hamiltonian}

We define the trace Hamiltonian by summing over all independent velocities:
\begin{equation}
\cH
=
\Tr\!\left(
\Pi_B \dot Q_B + \Pi_F \dot Q_F
+ \Pi_B^\dagger \dot Q_B^\dagger + \Pi_F^\dagger \dot Q_F^\dagger
\right)
-\cL.
\label{eq:Hdef}
\end{equation}
Substituting \eqref{eq:PiB}--\eqref{eq:PiFdag} and using \eqref{eq:Q1Q2defs}, one finds
\begin{align}
\Tr(\Pi_B \dot Q_B + \Pi_F \dot Q_F)
&= c\,\Tr\!\left(\lambda\,\dot Q_1^\dagger (\dot Q_B+\lambda\beta_2\dot Q_F)\right)
= c\,\Tr\!\left(\lambda\,\dot Q_1^\dagger \dot Q_2\right)
= \cL,
\\
\Tr(\Pi_B^\dagger \dot Q_B^\dagger + \Pi_F^\dagger \dot Q_F^\dagger)
&= c\,\Tr\!\left(\lambda\,(\dot Q_B^\dagger+\lambda\beta_1\dot Q_F^\dagger)\dot Q_2\right)
= c\,\Tr\!\left(\lambda\,\dot Q_1^\dagger \dot Q_2\right)
= \cL,
\end{align}
where in the second line we used that $\dot Q_2$ is bosonic and applied graded cyclicity.
Therefore
\begin{equation}
\boxed{
\cH = \cL
=
\frac{\hbar}{2\tauP}\;
\Tr\!\left[\lambda\,\dot Q_1^\dagger \dot Q_2\right].
}
\label{eq:H=L}
\end{equation}
This is the trace-dynamics analog of the Bateman-type cross-kinetic structure:
the Hamiltonian equals the Lagrangian, but need not be self-adjoint because
$\dot Q_1$ and $\dot Q_2$ are inequivalent.

\section{Bosonic vs fermionic contributions and the anti-self-adjoint part}

\subsection{Decomposition of the Hamiltonian}

Expanding \eqref{eq:H=L} using \eqref{eq:Q1Q2defs} gives
\begin{equation}
\cH = \cH_{BB}+\cH_{BF}+\cH_{FF},
\label{eq:Hdecomp}
\end{equation}
where
\begin{align}
\cH_{BB}
&=
c\,\Tr\!\left(\lambda\,\dot Q_B^\dagger \dot Q_B\right),
\label{eq:HBB}
\\
\cH_{BF}
&=
c\,\Tr\!\left(
\lambda^2\,\dot Q_B^\dagger \beta_2 \dot Q_F
+
\lambda^2\,\beta_1 \dot Q_F^\dagger \dot Q_B
\right),
\label{eq:HBF}
\\
\cH_{FF}
&=
c\,\Tr\!\left(\lambda^3\,\beta_1 \dot Q_F^\dagger \beta_2 \dot Q_F\right).
\label{eq:HFF}
\end{align}

\subsection{Adjoint assumptions for \texorpdfstring{$\beta_1,\beta_2$}{beta1,beta2}}

We now impose the following adjoint properties:
\begin{equation}
\boxed{
\beta_1^\dagger = -\beta_1,\qquad \beta_2^\dagger = -\beta_2,
\qquad \beta_1\neq\beta_2,\qquad \gr(\beta_1)=\gr(\beta_2)=1.
}
\label{eq:betaadj}
\end{equation}
We also assume the natural graded (anti)commutation with dynamical variables:
$\beta_a$ commutes with bosonic variables and anticommutes with fermionic variables.

\subsection{Self-adjointness of the purely bosonic sector}

From \eqref{eq:HBB} and \eqref{eq:adjoint},
\begin{equation}
\cH_{BB}^\dagger
=
c\,\Tr\!\left(\lambda\,(\dot Q_B^\dagger \dot Q_B)^\dagger\right)
=
c\,\Tr\!\left(\lambda\,\dot Q_B^\dagger \dot Q_B\right)
=
\cH_{BB}.
\label{eq:HBBsa}
\end{equation}
Hence the bosonic subsector has a self-adjoint Hamiltonian and does not by itself
supply an ASA contribution.

\subsection{An intrinsic ASA fermionic contribution}

Consider the fermionic term \eqref{eq:HFF}. Using \eqref{eq:adjoint} and \eqref{eq:betaadj},
\begin{align}
\cH_{FF}^\dagger
&=
c\,\Tr\!\left(
(\lambda^3 \beta_1 \dot Q_F^\dagger \beta_2 \dot Q_F)^\dagger
\right)
=
c\,\Tr\!\left(
\lambda^3\,\dot Q_F^\dagger \beta_2^\dagger \dot Q_F \beta_1^\dagger
\right)
\nonumber\\
&=
c\,\Tr\!\left(
\lambda^3\,\dot Q_F^\dagger (-\beta_2)\dot Q_F (-\beta_1)
\right)
=
c\,\Tr\!\left(
\lambda^3\,\dot Q_F^\dagger \beta_2\dot Q_F \beta_1
\right).
\label{eq:HFFdag1}
\end{align}
Now apply graded cyclicity \eqref{eq:gradedcyclic} to cyclically shift the final odd factor
$\beta_1$ to the front. Since $\beta_1$ is odd and the product
$\dot Q_F^\dagger\beta_2\dot Q_F$ is also odd (three odd factors), we obtain a minus sign:
\begin{equation}
\Tr(\dot Q_F^\dagger \beta_2\dot Q_F \beta_1)
=
-\,\Tr(\beta_1 \dot Q_F^\dagger \beta_2\dot Q_F).
\label{eq:cyclicsign}
\end{equation}
Combining \eqref{eq:HFFdag1} and \eqref{eq:cyclicsign} yields
\begin{equation}
\boxed{
\cH_{FF}^\dagger = -\,\cH_{FF}.
}
\label{eq:HFFasa}
\end{equation}
Thus $\cH_{FF}$ is \emph{purely anti-self-adjoint}. In particular, it provides an explicit
ASA contribution to the Hamiltonian, and this contribution vanishes identically when the
fermionic sector is absent.

\paragraph{Role of $\beta_1\neq\beta_2$.}
If one attempted to set $\beta_1=\beta_2=\beta$, the fermionic contribution collapses:
because $\beta$ is odd, $\beta \dot Q_F^\dagger \beta = -\beta\beta \dot Q_F^\dagger = 0$,
so $\cH_{FF}$ would vanish. Hence the requirement $\beta_1\neq\beta_2$ is not only
dynamically necessary \cite{Maithresh:2019ntp}; it is also structurally responsible for a nontrivial ASA fermionic term.

\subsection{ASA part of the full Hamiltonian}

Define the self-adjoint and anti-self-adjoint parts of the trace Hamiltonian by
\begin{equation}
\Hsa := \frac12(\cH+\cH^\dagger),\qquad
\Hasa := \frac12(\cH-\cH^\dagger).
\end{equation}
From \eqref{eq:Hdecomp}, \eqref{eq:HBBsa}, and \eqref{eq:HFFasa},
\begin{equation}
\boxed{
\Hasa
=
\frac12(\cH_{BF}-\cH_{BF}^\dagger)
+
\cH_{FF}.
}
\label{eq:Hasa}
\end{equation}
Independently of the detailed adjoint properties of the mixed term $\cH_{BF}$,
the key point is that \emph{every} contribution to $\Hasa$ contains fermionic variables:
if $\dot Q_F=\dot Q_F^\dagger=0$, then $\Hasa=0$ identically.

\paragraph{Remark (adjoint convention for $\beta_{1,2}$).}
The anti-self-adjoint character of the purely fermionic contribution
\(\cH_{FF}=c\,\Tr\!\big(\lambda^3\,\beta_1\,\dot Q_F^\dagger\,\beta_2\,\dot Q_F\big)\)
does not rely on taking \(\beta_a^\dagger=-\beta_a\) individually.  More generally, assume
\[
\beta_a^\dagger=\eta_a\,\beta_a,\qquad \eta_a\in\{+1,-1\},\qquad a=1,2,
\]
with \(\gr(\beta_a)=1\) and the standard GTD involution \((AB)^\dagger=B^\dagger A^\dagger\).
Then
\begin{align*}
\cH_{FF}^\dagger
&=c\,\lambda^3\,\Tr\!\Big((\beta_1\dot Q_F^\dagger\beta_2\dot Q_F)^\dagger\Big)
=c\,\lambda^3\,\eta_1\eta_2\,\Tr\!\Big(\dot Q_F^\dagger\beta_2\dot Q_F\beta_1\Big) \\
&=-\,c\,\lambda^3\,\eta_1\eta_2\,\Tr\!\Big(\beta_1\dot Q_F^\dagger\beta_2\dot Q_F\Big)
=-(\eta_1\eta_2)\,\cH_{FF},
\end{align*}
where the minus sign arises from graded cyclicity when moving the odd factor \(\beta_1\)
past the odd product \(\dot Q_F^\dagger\beta_2\dot Q_F\).
Hence \(\cH_{FF}\) is purely anti-self-adjoint whenever \(\eta_1\eta_2=+1\)
(i.e.\ both \(\beta_1,\beta_2\) are self-adjoint, or both are anti-self-adjoint).
In particular, the choice \(\beta_1^\dagger=\beta_1\) and \(\beta_2^\dagger=\beta_2\) yields the same
ASA fermionic term (and thus the same collapse channel) as the choice \(\beta_a^\dagger=-\beta_a\).

\section{Interpretation for spontaneous localization}

In trace dynamics, coarse-graining around statistical equilibrium yields two distinct regimes:
(i) if the Hamiltonian is effectively self-adjoint, one recovers emergent unitary quantum dynamics;
(ii) if the anti-self-adjoint component is significant, coarse-graining yields effective nonunitary
(stochastic) dynamics with spontaneous localization \cite{KakadeSinghSingh2023SpontaneousLocalisation}.

The computation above provides a structural mechanism for ``fermion-only collapse'' in GTD:
\begin{itemize}
\item The purely bosonic Hamiltonian $\cH_{BB}$ is self-adjoint and thus does not generate ASA-driven
collapse by itself.
\item The fermionic sector carries an intrinsic anti-self-adjoint term $\cH_{FF}$, present precisely because
$\beta_1$ and $\beta_2$ are unequal odd Grassmann elements inserted to make the unified trace Lagrangian bosonic.
\end{itemize}

Therefore, in regimes where ASA effects drive localization, the fundamental collapse channel is
naturally associated with fermionic degrees of freedom. Bosonic fields can still become classical
\emph{indirectly}, because in the STM-atom picture an STM atom consists of a fermion together with its
associated bosonic fields, so localization of the fermionic degrees forces classicality of the
associated bosonic sector through correlations/entanglement, without requiring independent bosonic collapse.

\section{Conclusions}

Starting from the unified STM-atom GTD trace Lagrangian, we computed the trace Hamiltonian via
trace-derivative canonical momenta, with bosonic and fermionic variations treated separately.
We found:
\begin{enumerate}
\item The trace Hamiltonian equals the trace Lagrangian (Bateman-type cross-kinetic structure),
but it need not be self-adjoint because $Q_1$ and $Q_2$ are inequivalent.
\item The purely bosonic contribution $\cH_{BB}$ is self-adjoint.
\item Assuming natural adjoint properties for the odd Grassmann elements $\beta_1,\beta_2$,
the fermionic contribution $\cH_{FF}$ is purely anti-self-adjoint, providing an explicit ASA component
of the Hamiltonian that vanishes in the bosonic subsector.
\end{enumerate}
This provides a first-principles explanation, within GTD, for why spontaneous localization acts
fundamentally on fermionic degrees of freedom and not on bosonic degrees of freedom.

\paragraph{Outlook.}
The next steps are to connect the magnitude of $\cH_{FF}$ (and fluctuations about equilibrium) to
collapse phenomenology (rates, noise kernels, and effective CSL parameters), and to embed the analysis
into the multi-STM-atom setting where entanglement and interactions become operative.

\bibliographystyle{unsrt}

\bibliography{octonion_bibliography_merged_updated, tracedynamics}

\end{document}